# ANALYSIS AND SYNTHESIS OF A SUBSYSTEM OF THE MANUAL CONTROL LOOP FOR MANNED SPACECRAFT

Yury Tyapchenko
Zhukovsky, Russia

## Abstract

A salient feature of the manned spacecraft is the predominance of discrete information in the manual control loop of the onboard systems. Specifically, command-signaling control panels (CSCP) as a subsystem of the manual control loop are widely used in the Russian manned spacecraft. In this paper CSCP are classified into four types: a) control panels based on multi-channel control; b) control panels based on command-information compression; c) control panels based on command and signaling information compression; d) integrated control consoles (ICC) based on computer and information technology. It is shown that ICC underlies modern information display systems (IDS). ICC first appeared in the Russian manned space program in the IDS of the Soyuz-TMA spacecraft and the Russian modules of the International Space Station. Results of engineering and psychological studies of different types of panels are produced.

## Analysis

It is known that Russian manned spacecraft are distinguished by a high level of automation of onboard control systems. This is especially becomes apparent in predominance of discrete data sent to the manual control loop of the spacecraft.

The manual control loop of the spacecraft is designed for execution and verification of commands issued by a cosmonaut as well as of the automatic and remote (from the ground) control commands.

In the manual control loop of the spacecraft, the so-called command-signaling control panels (CSCP) are used. CSCP is an aggregate of means for issuing discrete commands and for displaying discrete, usually on-off, signals.

The main elements of such CSCP are the information field (IF) and the command field (CF).

More than a hundred CSCP were created within the manned spaceflight program in the Soviet Union. They can be divided into four generation. Examples of control panels of each generation are shown in Figs. 1-4.

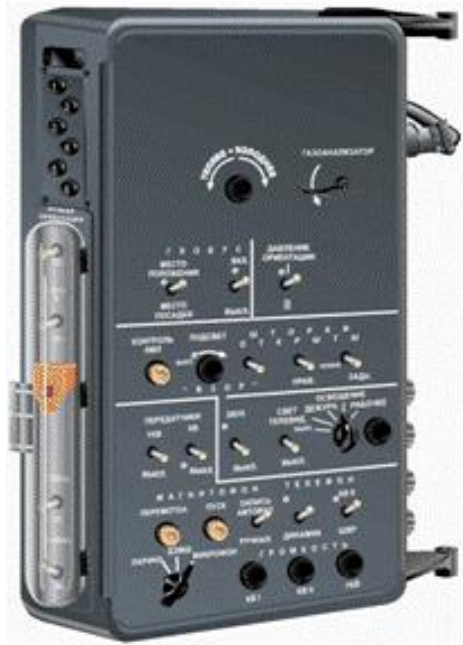

Figure 1. Multi-channel control panel.

Control panels of the first generation are based on multi-channel control (Figure 1). Control panels of subsequent generations are based on command and signaling information compression.

The second such control panel was of the command and signaling field (CSF) type, initially built for the 3KV-Number-6 spacecraft (see Figure 2).

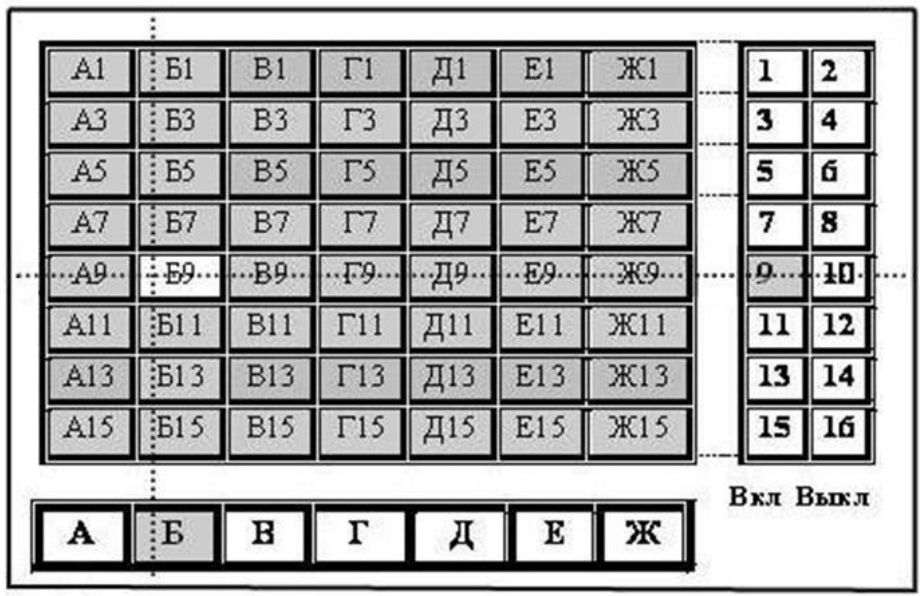

Figure 2. Control panel with a matrix method of issuing commands and an expanded information field.

Historically, the first control panel with information compression was of the command and signaling device (CSD) type, shown in Figure 3. This control panel was created within the lunar program and the program Soyuz-7K.

Within the space program, more than 10 CSD and more than 40 CSF were created.

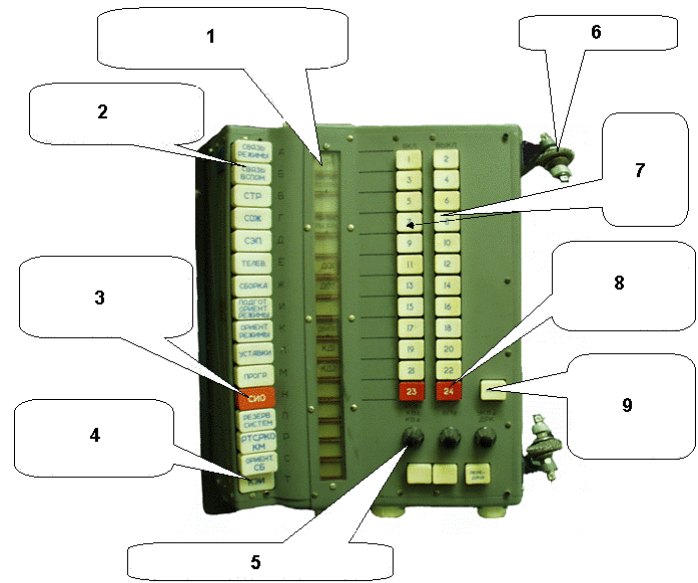

Figure 3. Control panel with a matrix method of issuing commands and a matrix method of selecting signals for verification:
1 – a stencil with names of commands, 2 – the keypad of systems selection,
3 – a safety important system, 4 – a system chosen for control, 5 – volume control,
6 – a shock-absorber, 7 – the keypad of command generating, 8 – safety important commands, 9 – lamp test.

CSF utilizes a matrix method of control (command-information compression) and an expanded method of a signaling information display. CSD utilizes both a matrix method of control and a matrix method of a signaling information display.

CSD is essentially an electromechanical display, in which the display formats are represented by a set of name strings for the controlled units. The name strings are placed on a cylinder in the direction of its axis. They are illuminated by a row of signal indicators placed inside a cylinder.

The next stage of CSD is the control panels based on onboard computers and monitors. The examples (see Figure 4) of such control panel are an integrated control console (ICC) of manual control loop of the Russian segment of the International Space Station, and a similar in structure integrated control console in the Neptune IDS for the Soyuz-TMA spaceship.

Before the transition to ICC, it became necessary to pose and solve the synthesis problem for CSCP. A timely solution of this problem in the manned spaceflight program permitted a drastic reduction of cost for developing a multitude of CSCP, successful minimization of the mass and sizes of control panels, ways for unification of components, etc.

Below I briefly outline the solution for the problem of the CSCP synthesis. It should be noted, that presently this work is not only of theoretical but also of practical interest, since all modern SDI are based on compression of commands and information.

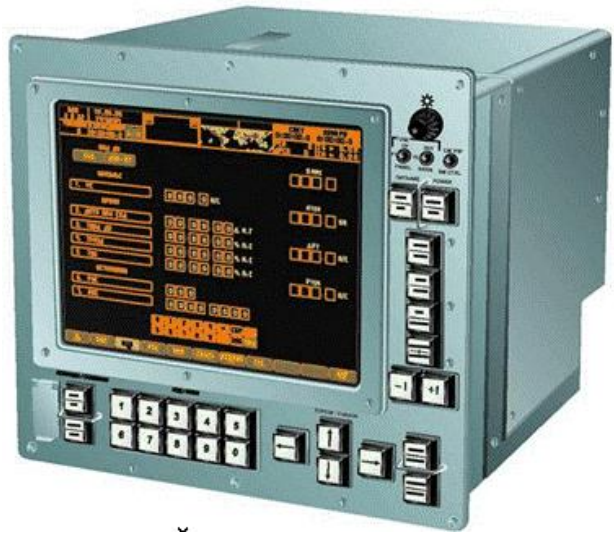

Figure 4. Integrated control console of a virtual command and signal matrix type, with a dialog system of interaction between an operator and the onboard system, a hierarchical system of displaying information about the unit, processes and the environment.

## The general approach to the synthesis of CSCP

As shown in Figure 5, all CSCP can be represented as a scale, on one side of which is a control panel with one signal indicator and one control, and on the other side is control panel containing many signal indicators and many controls, such as buttons or toggle switches.

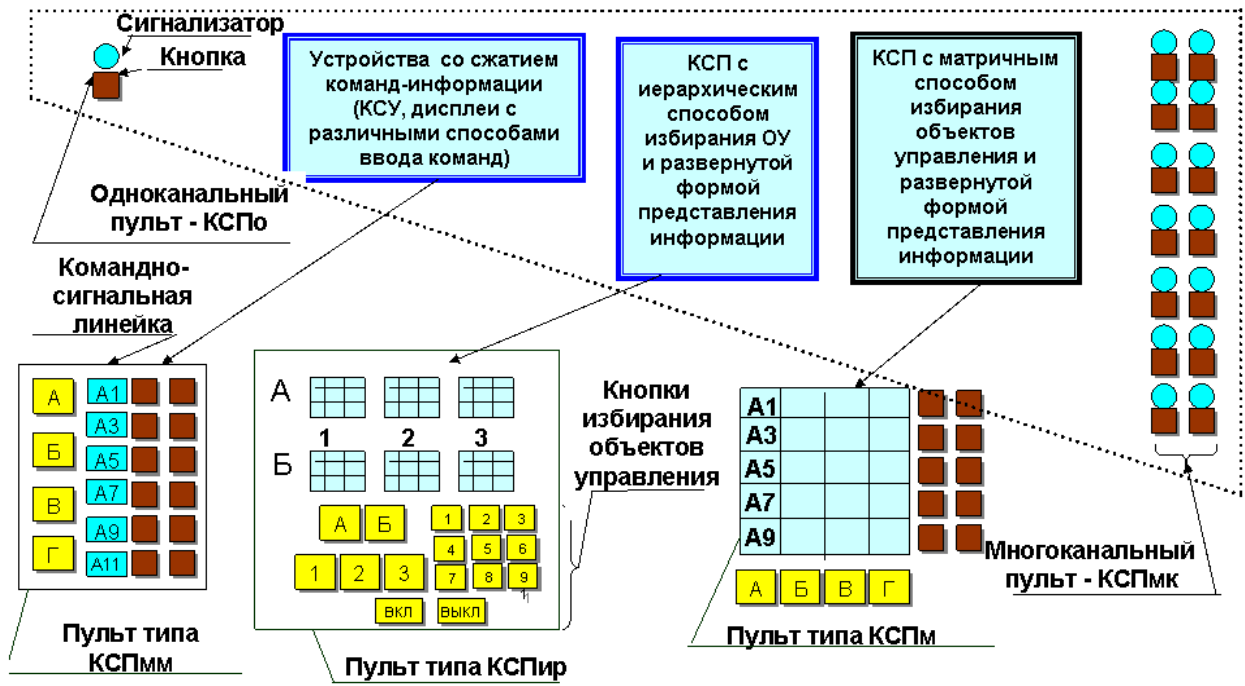

Figure 5. Scale of command-signaling control panels (CSCP).

The first control panel is a single-channel CSCP, conventionally speaking, a control panel with one light bulb and one button. The commands there are transmitted as a sequential code. The fact of the command execution is controlled by the bulb being lit up. This is a control panel with compression of signal and command-information. The degree of compression is described by coefficients K, one of which is equal to the ratio of the number of control units and the number of indicators on the control panel; another one is equal to the ratio of issued commands and the number of controls on the control panel.

The last control panel, on the opposite side, is a CSCP with multichannel information fields and command fields. This is a control panel with an expanded form of signal information and the number of controls equal to the number of controlled units. This is a control panel without compression of the information fields and command fields.

Obviously, the first control panel has the highest ratio of compression, and the last control panel has the lowest ratio of compression. Ratios of compression for the rest of CSPC are between these two extreme values.

**The objective of synthesis is the choice of a control panel within the range of compression ratios ensuring effective work of an operator and under the constraints of mass, size, reliability, power consumption, and other metrics.**

It is obvious that the control panels of the two extreme types cannot be used in IDS of big systems. The first CSCP require too long time to issue commands. The last CSCP require too large size and mass and also are unserviceable under g-overload.

Thus real life control panels can be found between the two extremes, and they must use some degree of command-information or signal information or both. For simplicity, we will call devices implementing command or signal information as "command-information compression devices".

In the above scale we highlight control panels with matrix and hierarchical methods for selecting controlled units:
- CSCPm = command-signal control panel with matrix command issuing;
- CSCPmm = command-signal control panel with matrix command issuing and matrix choice of controlled units.

These are control panels with a two-stage selection of controlled units. Two-stage selection is a special case of multi-stage, or hierarchical, method of selection controlled units.

An example of a control panel with a hierarchical method of selection and with a spatial separation of command and information fields (CSCPis) is shown in the same figure.

The set of CSCPir can be broken into three sub-sets:
- CSCPirsc = control panels with a spatial combination of IF and CF;
- CSCPirss = control panels with spatial separation of IF and CF;
- CSCPint = control panels with a combination of IF and CF in the same device, i.e. CSCP of integrated type.

In the following we present in sequence the solutions for synthesis of CSCPm, CSCPmm, and CSCPir. We adopt the following criteria of optimization:
- minimum of controls, translating to the mass and the sizes of a control panel;

- time for issuing commands;
- subjective evaluation by operators.

## Synthesis of CSCPm

A typical structure of CSCPm with spatial combination of a keyboard and an information field is shown in Figure 2.

Button switches for selection of systems (buttons with letter symbols on them) remain in a fixed state after pressing. Buttons for issuing commands react to every push.

For the minimization of the number of controls, control panels with matrix method of control (such as CSCPm and CSCPmm) are optimal, provided that the control matrix, i.e. a decoder of issued commands, is square. This corresponds to the number of buttons for system selection being equal to the number of buttons for issuing commands.

If controlled units are two-state, the buttons for issuing commands are grouped in twos, as shown in Figure 2. Thus the information field becomes rectangular with the aspect ratio of 2.

To repeat, the optimal CSCPm, relative to the criterion of minimal number of controls, is a control panel with a square control matrix.

In practice, more often CSCPm with a square IF and various placements of buttons relative to IF are used.

Layouts with button switches on the left and below IF were used in control panels for work with left hand. Layouts with button switches on the right hand below IF were used in control panels for work with right hand. Layouts with button switches above IF have not been used.

CSCPm with separated keyboard and IF do not comply with ergonomic requirements and have not been used in practice.

## Synthesis CSCPmm

An example of a control panel CSCPmm is shown in Figure 3. This is a control panel of the CSD type for the Sirius IDS in the spaceship Soyuz-7K [1].

Button switches for selection of systems (buttons with letter symbols on them) remain in a fixed state after pressing. Buttons for issuing commands react to every push.

First, a system is selected by pressing a letter button. Strings with the names of controlled units appear above a row of indicators. The name decoder in the Sirius IDS is implemented as a lace multi-face cylinder. The number of faces is equal to the number of controlled systems. Transparent templates with the names of units are inserted in the faces. Turning a unit on or off is accomplished by pressing the command button. If a unit is on, the corresponding indicator bulb within a row of indicators is lit, and thus the template in front of it is illuminated.

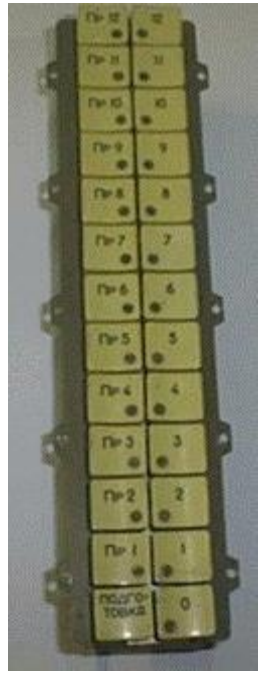

Figure 6. CSCPmm without a command name decoder.

To monitor the program modes of control in IDS, CSCPmm without name decoders were widely used. An example of such control panel is shown in Figure 6.

As a rule, CSCPmm of this type contains two groups of buttons: one with a fixed state for selection of programs, and the other one for issuing commands (within the selected program) by pressing them.

Each button contains an LED. Turning a program on and off is monitored by lighting of an LED in corresponding program button. After an assigned time from the start of the program, control commands are being issued. If a command succeeded, and LED in the second row lights up. If a command did not issue in an assigned time, an operator manually issues it.

This demonstrates how a simple device based on command-information compression successfully solves the task of redundancy of the program circuit of a control system.

According to the criterion of minimal number of controls, control panels of the CSCPmm type are optimal, as it was shown before, when the matrix of command decoding is square. If the controlled units are two-state, the command-issuing buttons are grouped by two, just like in CSCPm. In practice, mostly CSCPmm with square on near-square command decoder are used. In this case, the command compression coefficients in CSCPm and CSCPmm are the same.

The creation of CSCPmm type in the 1960s (i.e. CSCP of the Sirius IDS for the Soyuz-7K spacecraft) opened a pathway for integration of control and display means. It led to creation of IDS based on personal computers.

**Synthesis of CSCPir**

As the number of controlled units increases, the size of CSCPm increases, and a part of control means becomes out of reach for an operator.

The increased number of commands in the Salyut space station, and the predominance of control functions in the cosmonauts' activity posed the problem of spatial separation of IF and CF and the problem of optimization of the hierarchical (multi-stage) method of control unit selection. Practically, these problem were solved in the Mirzam IDS for the Salyut-17K space station and in the Pluton IDS for the Mir space station [2].

According to the criterion of minimal number of controls, a control panel with the compression coefficient of three is optimal. It is interesting to note, that the use of any IF based on a 3x3 array is optimal. It corresponds to the decimal keypad with 3×3 layout and numbers from1 to 9 on its keys, as shown in Figure 7.

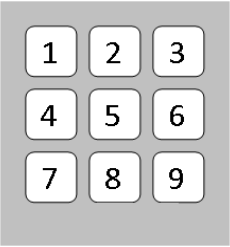

Figure 7. Button switch for the numeric keypad of 3×3 layout without the digit zero.

The selection of an array is performed in a hierarchical scheme. In practice, the number of stages is chosen as small as possible, since a decreased number of stages leads to an insignificant increase of the number of controls, but significantly increases the selection time.

## Comparison of control panel specs

The values for mass, the number of external wires, the area of the monitor, and the required power are shown in Figure 8 for a multi-channel CSCPmc, CSCPm with an expanded IF, and the CSCPmm in the Soyuz-7K spacecraft. The values for CSCPmm are set as unity. The values for other CSCP are calculated. This shows that the specs of CSCPmm are significantly better than those of other CSCP.

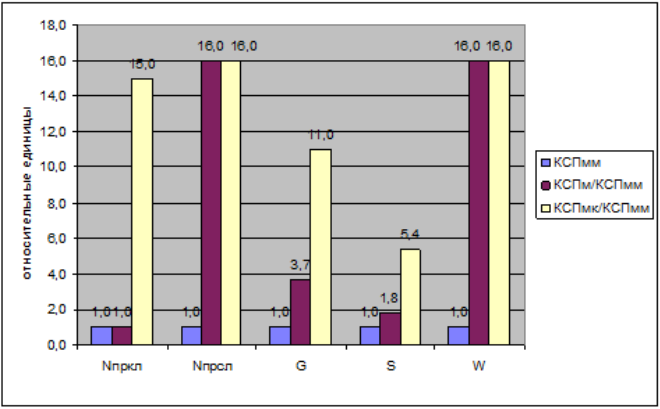

Figure 8. Comparison of CSCP characteristics, where Nprkl and Nprsl are the numbers of wires for command and signaling rows, respectively, G is the mass, S is the area, W is the power consumption.

Figure 9 shows the operator response time for solving the same tasks using CSCP of different types: CSCPm, CSCPmm, and CSCPmc (a multi-channel control panel using toggle switches) [3]. The following conclusions are made from these results:
- there is no significant difference in response time between CSCPm and CSCPmm;
- in control systems, in which the units are well classifiable by well-distinguishable features, control panels with matrix selection and expanded/contracted information display can be used.

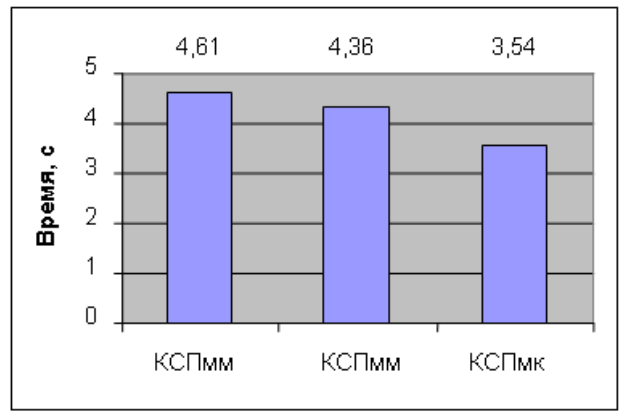

Figure 9. Operator response time while working with control panels.

Despite the obvious advantages of their specs, CSCPmm were not developed further and were replaced with CSCPm (i.e., CSCP with expanded information display) when the Soyuz-T and Soyuz TM spacecraft, as well as Mir space station were built.

Considering that a return to expanded information displays was technologically not a promising endeavor, the author led research in two directions:

a) The first and major direction: the study of the causes of negative evaluations of control panels with command-information compression; the search for ways to display compressed information.

b) The second and secondary direction, which conceptually did not have a long term perspective, but for many years required most of the effort: the development of the structure of IF, identification of methods and means for issuing a large number of commands, and the search for methods to increase the efficiency cosmonauts' work on large arrays of signal information in CSCP with an extended display.

In the first direction, we carried out:
- The studies of operator activities during the phase of complex ground tests of IDS Sirius-7K for the spacecraft Soyuz-7K.
- Research on the reasons for of preference of CSCPm to CSCPmm.

In the second direction, we carried out comparative studies of the operator efficiency while working with control panels with expanded information display and various methods of command issuing. Recommendations for increase the operator work efficiency while working with a large volume of signaling information have been developed.

## The first direction

The work of operators of the descent module of the Soyuz-7K spacecraft was studied on the stage of ground complex tests at the manufacturing factory. Also the work of cosmonauts on the simulator was studied during their training for the mission at the Gagarin Cosmonaut Training Center.

It is well known that the Soyuz-7K spacecraft contains two control panels of the CSCPmm type – the left CSDl and the right CSDr. Control of systems can be executed from CSDl, from CSDr, or from both simultaneously. Each CSD contains 16 buttons for system selection and 12×2 buttons for issuing commands. Each system has up to 16 controlled units. Monitoring of state of systems is performed after selecting a system. Issuing of commands is done in two stages: selection of the system and then pressing the "on" or "off" buttons.

One can classify the following operations in using CSD:
- K-operation – control of the state of the units by the operator;
- U-operation – an operation resulting in the operator's issuing a command to turn an unit on or off;
- L-loop or R-loop – a consecutive call by the operator of systems to check the status of units in each system using CSDl and CSDr, respectively;
- O-loop – a consecutive pressing by the operator of all system selection buttons on CSDl or CSDr to check whether units are on or off units after passing a program label or after a command via the radio channel from the ground.

The studies on the stage of ground complex tests showed:
- Compression of command-information adopted in CSD of Soyuz 7K spacecraft leads to an increase of operator load during checking runs by 1.2 to 1.3 times, and in automatic mode by 2 to 3 times compared to a control panel of CSCPm type with an equivalent information capacity. Compared to a conventional control panel, these values are 1.6 and 4.5 times, respectively;
- O-loops, i.e., cyclical control operations constituted 61% of all CSDr operation. When using CSCPm and CSCPmc, such operations are absent.

Research performed in the simulator TDK-7K at the Gagarin Cosmonaut Training Center in 1970 and 1973 confirmed the above findings, both quantitatively and qualitatively. It is worth noting that the number of K-operations performed by cosmonauts who have been training in 1971 and 1972, decreased in 1973. This suggests that as skills develop, the trust of control panels with compressed command-information increases.

Thus the use of compression of command-information leads in principle leads to a significant increase in the load of operators during monitoring of automatic control modes.

The above findings do not address the reasons for preference of CSCP with an expanded IF to the CSCP with contracted IF.

We put forward a hypothesis that seemingly non-essential differences in ergonomic qualities of CSCPm and CSCPmm control panels on the device level become essential on the system level. Considering identical information provided to the operators by CSCPm and CSCPmm control panels, the reasons for this preference must be psychological.

Research in Ref. [4] showed that the main factor by which CSCPm are preferred to CSCPmm is the lack of relevant information provided to an operator by CSCPmm. CSCPm were more familiar to operators and training to work with CSCPm was perceived as easier by the operators. This created a bias among operators that CSCPm were better than CSCPmm. The two factors (information relevance and operators' bias) are closely related, reinforcing each other.

From the results of this research, the main recommendation for building control panels with compression of command-information was:
**Layout of the information field should satisfy the main principle - the principle of autonomy.**

This means that the display of status of the units of a given system should be placed in one row of CSCPmm; or one row of CSCPmm must address one task. If in layout of IF by the functional principle, the same units take part in several functions, the state of these units must be displayed in each row, from which control is executed.

To reduce the operator load in the control of automatic modes, the author proposed the principle of signaling changes in the status of controlled units.

**The principle of autonomy and the principle of signaling status changes became the fundamental principles for building modern computerized SDI.**

## The second direction

A possibility to build control panels with spatial separation of IF and CF investigated by F. E. Temnikova and B. Panshin [5].

In our work [6], three ways to control unit selection were compared: multi-stage way (control panel CSCPi), address way (control panel CSCPa with a numeric keypad for selecting address of a controlled unit) and the matrix way. Last way considered in two options: combination of IF and CF (control panels CSCPm) and spatial separation of IF and CF (control panel CSCPmr). The main results of the study are shown in in Figure 10.

The following conclusions and recommendations were made from the results of this study:

- Of the two control panels with a matrix selection, CSCPm with keypads along perpendicular edges of the signal field are preferable. In those, CSCPm with the field size of 10×10 cells are close to optimal;

- In random selection of controlled units, efficiency of multi-stage control panels (CSCPi) depends on breaking IF into parts. The limit of breaking is two stages, i.e. the matrix way of selection. The efficiency of CSCPi can be increased for tasks solved consecutively by the operator. In this case, for example, a system and a corresponding screen are selected first, and then work is done within the IF of the selected screen;

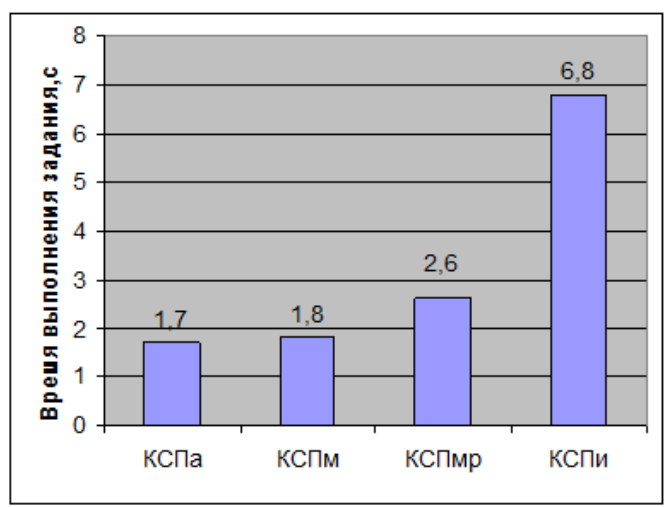

Figure 10. Average time of task completion while working with control panels CSCPa, CSCPm, CSCPmr, and CSCPi.

- The address method of coding is not worse than the matrix method in efficiency. However in practice, it makes coding all cells of a signal screen necessary. This decreases the area of cells available for meaningful labels. It goes against the ergonomic requirements and size requirements, as the screen size grows.

It is beneficial to separate the control task in two stages: a) selection of the screen by the address method; b) selection of a cell within the screen.

Just like in the synthesis of a control panel with hierarchical selection, a screen with a size of 3×3 with a corresponding decimal keyboard of similar structure (see Figure 7) is recommended from the results of engineering psychology studies.

Such a control panel not only satisfies ergonomic requirements, but also the requirements of display unification and control means unification. It complies with the trend of transitioning control panels to computer display hardware.

Based on the recommendations of these studies, the IDS Pluton and Mercury for the Mir space station were built and a number of prototypes of control panels for mobile power plants were created. At the same time, the modes of unit status control and the dark screen mode were introduced. Operation of IDS Pluton in the Mir space station with these modes confirmed their high efficiency.

**Conclusion**

In the above, the evolution of hardware of control panels was presented. In this development, CSCPi with the above structure of the screen were the final type of control panels with an expanded IF. The hardware path of development of such control panels is exhausted. Future control panels are computerized ones with contracted IF based on modern computer and information technologies. However their design has to satisfy requirements, many of which were defined from the results of research presented in this work.

## Acknowledgement

The author is grateful to Dmitri E. Nikonov for translating the article from Russian.